\documentclass{preprint}

\usepackage{latexsym}
\usepackage{graphicx}
\usepackage{multicol,multirow}
\usepackage{amsmath,amssymb,amsfonts}
\usepackage{mathrsfs}
\usepackage{amsthm}
\usepackage{apacite}
\usepackage{rotating}
\usepackage{appendix}
\usepackage[authoryear]{natbib}
\usepackage{ifpdf}
\usepackage[T1]{fontenc}
\usepackage{times}
\usepackage{sourcesanspro}
\usepackage{newtxmath}
\usepackage{textcomp}%
\usepackage{xcolor}%
\usepackage{hyperref}
\usepackage[section]{placeins}
\usepackage{hyperref}

\articletype{RESEARCH ARTICLE}
\jname{Data-Centric Engineering}
\jyear{2020}
\jdoi{10.1017/dce.2020.7}

\begin{document}

\begin{Frontmatter}

\title[]
{Wind speed inference from environmental flow-structure interactions}

\author[1]{Jennifer L. Cardona}\orcid{0000-0003-3491-0638}
\author[2]{Katherine L. Bouman}\orcid{0000-0003-0077-4367}
\author*[3]{John O. Dabiri}\email{jodabiri@caltech.edu}


\address[1]{\orgdiv{Department of Mechanical Engineering}, \orgname{Stanford University}, \orgaddress{\street{Stanford}, \state{California}, \postcode{94305}, \country{USA}}}

\address*[2]{\orgdiv{Computing and Mathematical Sciences \& Electrical Engineering \& Astronomy}, \orgname{California Institute of Technology}, \orgaddress{\street{Pasadena}, \state{California}, \postcode{91125}, \country{USA}}}

\address*[3]{\orgdiv{Graduate Aerospace Laboratories \& Mechanical Engineering}, \orgname{California Institute of Technology}, \orgaddress{\street{Pasadena}, \state{California}, \postcode{91125}, \country{USA}}}



\abstract{This study aims to leverage the relationship between fluid dynamic loading and resulting structural deformation to infer the incident flow speed from measurements of time-dependent structure kinematics. Wind tunnel studies are performed on cantilevered cylinders and trees. Tip deflections of the wind-loaded structures are captured in time series data, and a physical model of the relationship between force and deflection is applied to calculate the instantaneous wind speed normalized with respect to a known reference wind speed. Wind speeds inferred from visual measurements showed consistent agreement with ground truth anemometer measurements for different cylinder and tree configurations. These results suggest an approach for non-intrusive, quantitative flow velocimetry that eliminates the need to directly visualize or instrument the flow itself.}

\begin{policy}[Impact Statement]
{We present a velocimetry method that infers time-dependent flow speeds using visual observations of flow-structure interactions such as the swaying of trees. This can alleviate the need to directly instrument or visualize the flow to quantify its speed, instead relying on preexisting objects in an environment that are deflecting due to the incident flow. The method has the potential to turn ubiquitous objects like trees into abundant, natural, environmental flow sensors for applications such as weather forecasting, wind energy resource quantification, and studies of wildfire propagation.}

\end{policy}

\end{Frontmatter}

\section{Introduction}
\label{sec:intro}

Fluid-structure interactions such as the bending and swaying of trees in the wind provide visual cues that contain information about the surrounding flow. If visual measurements of deflections can be used to infer quantitative estimates of local wind speeds, then common objects like trees could be used as abundant natural anemometers, requiring only non-intrusive visual access to record wind speed measurements. This would potentially be useful in applications such as data assimilation for weather forecasting, wind energy resource quantification, and understanding wildfire behavior. Recent work has examined this visual anemometry task through a data-driven approach, where a neural network based model was trained to output wind speeds based on input videos of flags and trees in naturally occurring wind \citep{Cardona2019SeeingNetwork}. However, achieving a data-driven model that would generalize to a wide variety of objects (e.g.\ trees of different sizes and species) could potentially require an extensive data collection campaign. Physical models for fluid-structure interactions may be advantageous in providing a framework that could be used for visual anemometry across a broader range of structures. Here, we focus on objects that can be modelled as cantilever beams under wind loading.

Flow-sensing cantilevers are found in nature. For instance, the lateral line system allows fish to sense the surrounding flow via the deflection of hair-like structures \citep{Bleckmann2009LateralFish.}. Artificial lateral line sensors have been developed to mimic this flow sensing function as discussed by \citet{Shizhe2014UnderwaterSensors}. Cantilever beam deflections have also been used to measure wind speeds specifically. \citet{Tritton1959ExperimentsNumbers} used optical measurements of cantilevered quartz fiber deflections, and \citet{Kraitse1977MemorandumAnemometer} used strain gauge measurements of millimeter-scale silicone beams. These drag-based anemometers relied on knowledge of the material properties of the beam. These physical properties could then be used in conjunction with beam bending theory to calculate the drag force on the beam and quantify the wind speed. In these cases, the beam materials were specifically chosen for sensing purposes. Another example of cantilever deflection-based anemometry can be seen in \citet{Barth2005Laser-cantileverFlows}, where the deflection of a millimeter-scale cantilever affects the position of a reflected laser, which is calibrated to measure flow speeds. 

While these prior studies have shown success in measuring flow speeds by instrumenting the flow with cantilevers specifically designed and intended for sensing, the present work aims to extend the concept of flow-sensing cantilevers so that it can ultimately be used with a variety of pre-existing structures in an environment without further instrumenting the flow. Natural structures such as trees have material properties that are unknown \textit{a priori}, and they are more geometrically complex than single beams. Hence, in this work we exploit simplified models of the flow-structure interactions to avoid the need for direct consideration of these details, while still capturing the physical influence of the wind on structure deformation. By this approach, we can leverage the prevalence of trees and other vegetation in both rural and built environments for use in visual anemometry.

Approximate wind speed scales have been developed based on field observations of fluid-structure interactions in the past. The Fujita scale, for example, is used to infer tornado wind speed based on the damage to structures in its path \citep{Doswell2009OnUSA}. This has been particularly useful since more conventional wind speed measurements are rare and difficult to obtain for tornadoes. The Beaufort scale is another well-known wind speed scale that relies on visual cues. The version of the scale adapted for use on land employs qualitative descriptions of tree behavior (e.g.\ branch motion or breaking of twigs) to estimate an instantaneous wind speed range, and it has also been applied to region-specific vegetation \citep{Jemison1934BeaufortMountains}. Visual observations of trees and other vegetation have also been used to estimate mean annual wind speeds. The Griggs-Putnam Index uses qualitative descriptions of tree deformation to categorize mean annual wind speeds into seven binned increments of 1-2 $ms^{-1}$ \citep{Wade1979}. Time series measurements of tree deformations and a physical model for the fluid-structure interactions may allow for an extension to quantitative, instantaneous flow speed measurements. 

Wind-tree interactions have been widely studied \citep{deLangre2008EffectsPlants}. Prior investigations have used various forms of cantilever beam models to describe tree behavior. For instance, \citet{Kemper1968LargeBeams}, \citet{Morgan1987StructuralLoading}, and \citet{Gardiner1992MathematicalTrees} compared tree deflections to tapered cantilever beams. The relationship between wind speed and drag force on trees has also been explored, in particular with regard to the drag reduction that results from large deformations of the tree crown. Several studies have observed and quantified the drag on trees as a function of wind speed \citep{Fraser1962WindTrees, Mayhead1973, Rudnicki2004WindSpecies, Vollsinger2005WindSpecies, Kane2006DragMaple, Koizumi2010EvaluationMethod, deLangre2012OnPlants, Manickathan2018ComparativeTunnel}. Despite the complexities of wind-tree interactions, this prior work suggests that the structural behavior under wind loading can be generalized in physical models.

The present work aims to infer incident wind speed measurements from observations of cantilevered cylinders and trees. We achieve this visual anemometry without \textit{a priori} knowledge of the material properties of the structures, as is the case for application to natural vegetation. A model for the relationship between the drag force and mean deflection is proposed. Using this model, normalized wind speeds can be approximated based only on the measured deflections of a structure, where the normalization is based on the measured wind speed and deformation at a selected reference time. The model was tested and compared to ground truth anemometer-measured wind speeds for both cantilevered cylinders and trees in wind tunnel experiments. Given that object deflections can be observed from videos, this method can serve as a non-intrusive technique to measure normalized wind speeds, where dimensional velocities may be recovered with a single calibration measurement at a reference wind speed. This eliminates the need to directly instrument or visualize the flow for quantitative characterization.

\section{Physical Model}\label{sec:model}
A physical model was used to relate wind speed to deflections based on a force balance. The dynamic pressure, $p$, on an object in flow is proportional to the product of the fluid density, $\rho$, and the square of the mean incident wind speed, $U$ \citep{Batchelor2000AnDynamics}. The mean force of the wind, $F_W$, is given by multiplying $p$ by the projected frontal area, $A$, and is therefore proportional to $\rho U^2A$:

\begin{equation}
    F_W = pA\\
    \propto \rho U^2 A
    \label{eq:F_D}
\end{equation}

We model the deflection of the structure following Hooke's Law:

\begin{equation}
    F_E = \kappa\delta
    \label{eq:hooke}
\end{equation}

\noindent where $F_E$ is the elastic restoring force, $\kappa$ is the elastic constant, and $\delta$ is the deflection of the free end of the cantilever. This model is applicable for small deflections under point loads or distributed loads, with differences in the configuration of loading captured in the form of the constant of proportionality. For a cantilever of constant cross section subject to a uniformly distributed load, the relationship between force per unit length, $f$, and the tip deflection given by Euler-Bernoulli beam theory follows:

\begin{equation}
    f = \left( \frac{8EI}{L^4} \right )\delta
\end{equation}

\noindent where $E$ is the Young's modulus of the material, $I$ is the area moment of inertia, and $L$ is the length of the beam. In this case, the elastic constant is related to the geometric and material properties of the cantilever. The total force on the beam is equal to $fL$, and the elastic constant as defined in equation \ref{eq:hooke} is $\kappa = \frac{8EI}{L^3}$. 

A balance of the forces in equations \ref{eq:F_D} and \ref{eq:hooke} above gives a relationship between the incident flow speed and the structure deformation:

\begin{equation}
    U \propto \sqrt{\frac{\kappa\delta}{\rho A}}
    \label{eq:single_speed_relation}
\end{equation}

\noindent Furthermore, assuming that $\rho$, $\kappa$ and $A$ remain constant under the conditions of interest, the wind speed normalized by a non-zero reference condition characterized by $U_0$ and $\delta_0$ is given by:

\begin{equation}
    \frac{U}{U_0} = \sqrt{\frac{\delta}{\delta_0}}
    \label{eq:ratio}
\end{equation}

\noindent The normalized wind speed given in equation \ref{eq:ratio} is independent of the material properties of the structure, and the dimensional wind speed can be recovered given only a measurement of $\delta$ and the reference condition ($U_0$, $\delta_0$). In the measurements described below, we examine the regime of validity of the model in equation \ref{eq:ratio}.

\subsection{Modification for Tree Crown Deformation}
Experimental observations by \citet{Roodbaraky1994ExperimentalTrees} have shown that load-deflection curves for various tree species appear to be linear, which suggests that the linear relationship proposed in equation \ref{eq:hooke} is appropriate to use for trees. However, equation \ref{eq:ratio} assumes that the frontal area of the structure is constant for all incident wind conditions. Prior observations have shown a reduced growth in drag force on trees with increasing $U$ \citep{Fraser1962WindTrees, Mayhead1973, Kane2006DragMaple, Koizumi2010EvaluationMethod, deLangre2012OnPlants, Manickathan2018ComparativeTunnel}. The drag reduction is attributed to reconfiguration of the tree crown which leads to streamlining as a result of the change in area \citep{Harder2004ReconfigurationHabitats, Rudnicki2004WindSpecies, Vollsinger2005WindSpecies, Manickathan2018ComparativeTunnel}. The effect of the changing area has often been taken into account through the use of a Vogel number \citep{Vogel1989DragWinds}, $\beta$, and the drag is assumed to grow as $U^{2 + \beta}$, where $\beta$ has a negative value to compensate for the streamlining effect. The Vogel number has been found to vary between tree species, with typical magnitudes in the range of $\mathcal{O}([0.1, 1])$ \citep{deLangre2012OnPlants, Manickathan2018ComparativeTunnel}. In general, the value of $\beta$ may be unknown for a tree of interest without extensive testing. Therefore, instead of accounting for reconfiguration by using a Vogel exponent, the changing instantaneous frontal area was taken directly into account. Considering that $A$ changes with wind speed, the normalized wind speed becomes:

\begin{equation}
    \frac{U}{U_0} = \sqrt{\frac{\delta A_0}{\delta_0 A}}
\end{equation}

\noindent where $A_0$ is the frontal area of the tree at reference speed $U_0$. Where a projection of the frontal area of the structure is not available, the change in area can be approximated by assuming $A \propto h^2$, where $h$ is the projected tree height. This gives the modified form of the model to determine normalized wind speeds from tree deflections:

\begin{equation}
    \frac{U}{U_0} = \sqrt{\frac{\delta h_0^2}{\delta_0 h^2}}
    \label{eq:ratio_trees}
\end{equation}

\section{Experimental Methods}
\label{sec:methods}

\subsection{Cantilevered Cylinder Deformation Measurements}
\label{sec:methods_cylinder}

\begin{figure}[!ht]
    \centering
    \includegraphics[width = \linewidth, , trim=0 0 0 250, clip]{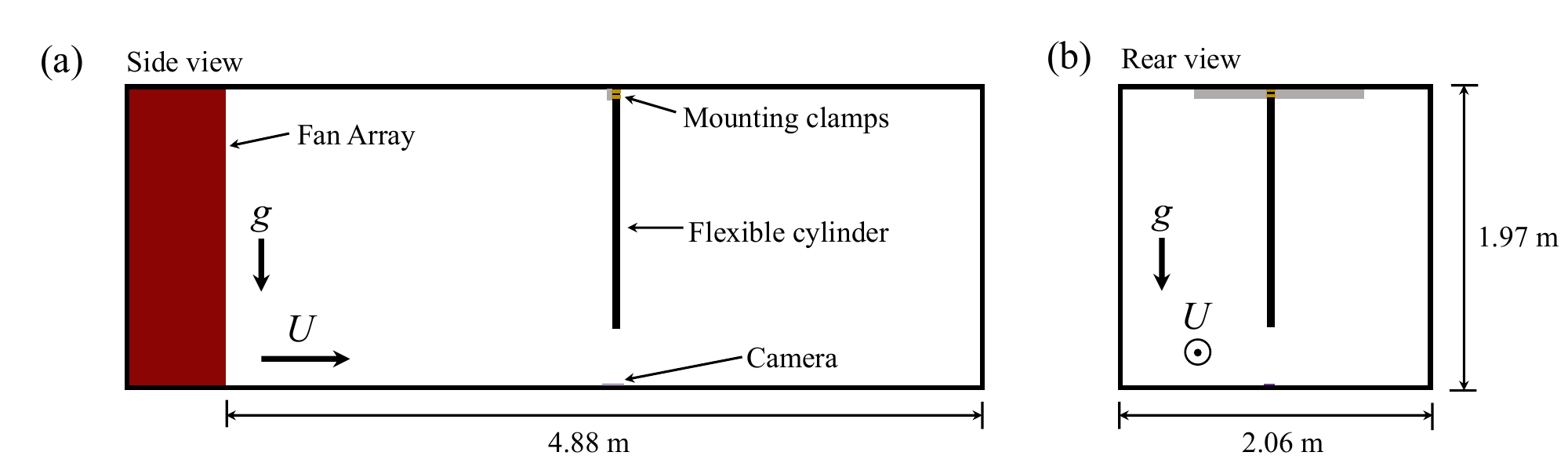}
    \caption{Schematic of experimental setup to measure cylinder deflection showing (a) side and (b) rear views. Directions of flow and gravity are indicated. Cylinder dimensions are to scale for the PVC tube of $D=$ 5.1 $\pm$ 0.1 cm, $L =$ 1.52 $\pm$ 0.01 m (the cylinder with the largest frontal area)}
    \label{fig:cylinder_setup}
\end{figure}

A first set of experiments studied deflections of flexible cylinders with circular cross-section. These canonical structures contrast with more complex tree geometries studied subsequently. Experiments were carried out in a 2.06 m $\times$ 1.97 m cross-section open-circuit wind tunnel (more details can be found in \citet{Brownstein2019AerodynamicallyFlow}). Mean flow speeds were measured using a factory-calibrated anemometer (Dwyer Series 641RM Air Velocity Transmitter) with a digital readout. The anemometer was accurate to $3\%$ of full scale, where full scale was set to 15 ms$^{-1}$. Each test cylinder was rigidly mounted to the ceiling of the tunnel, and subjected to three mean flow speeds ($U=$ [4.5, 5.6, 6.6] $\pm$ 0.5 ms$^{-1}$). The maximum blockage ratio in the wind tunnel based on the projected area of the test cylinder and mounting apparatus was $4.0\%$. A schematic of the setup is shown in figure \ref{fig:cylinder_setup}.

\begin{figure}[!hbt]
    \centering
    \includegraphics[width = 0.97\linewidth, , trim= 0 0 0 80, clip]{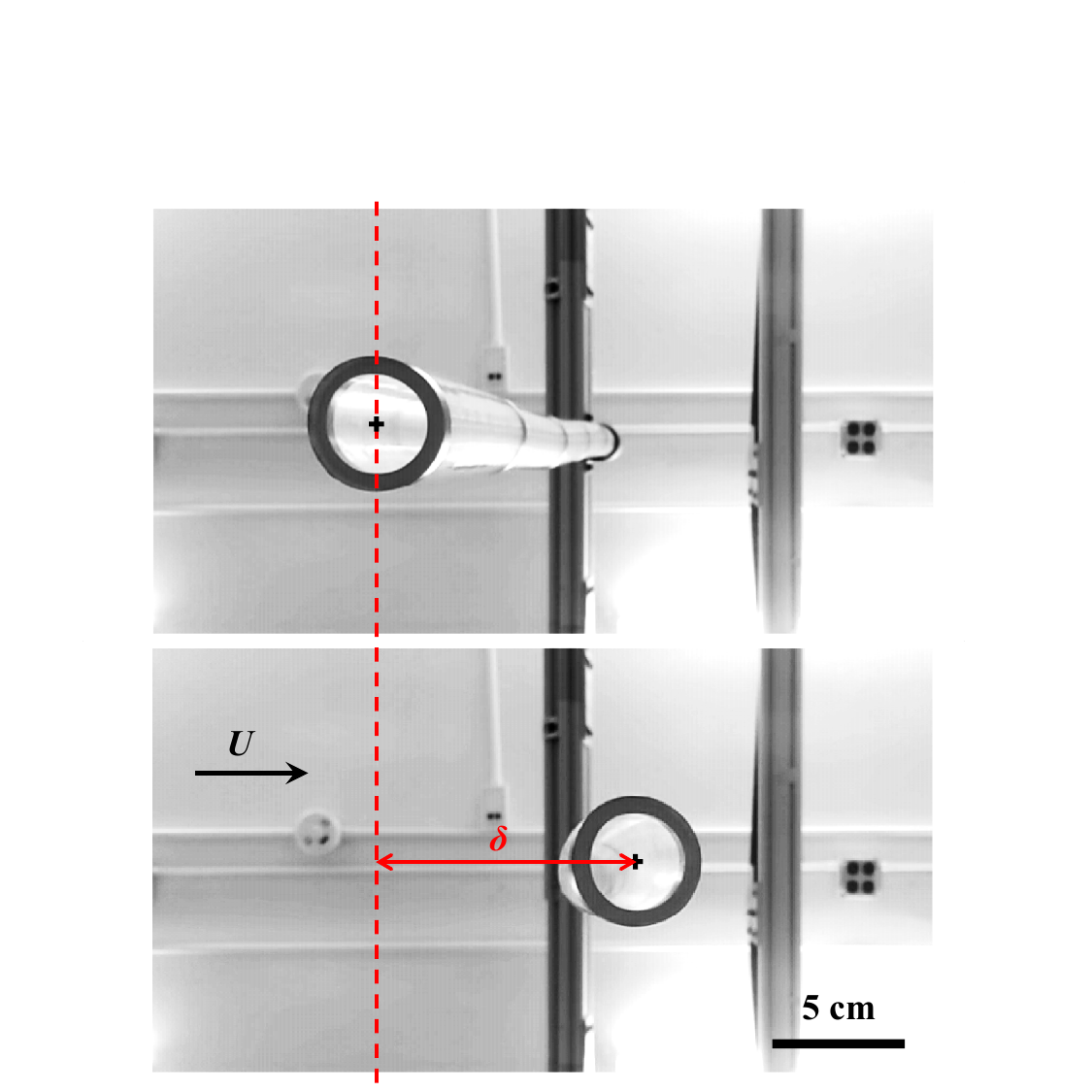}
    \caption{Representative frames showing the displacement of the cylinder free end (PVC tube, $D =$ 5.1 $\pm$ 0.1 cm). Top: cylinder surface under no wind load. Bottom: Displaced cylinder subject to incident flow speed $U =$ 5.6 $\pm$ 0.5 ms$^{-1}$, with streamwise displacement, $\delta$, shown in reference to center position under no load. Cylinder centers indicated with `+'}
    \label{fig:cylinder_displacement}
\end{figure}

\begin{figure}[!hbt]
    \centering
    \includegraphics[width = 0.9\linewidth, trim=0 300 0 310, clip]{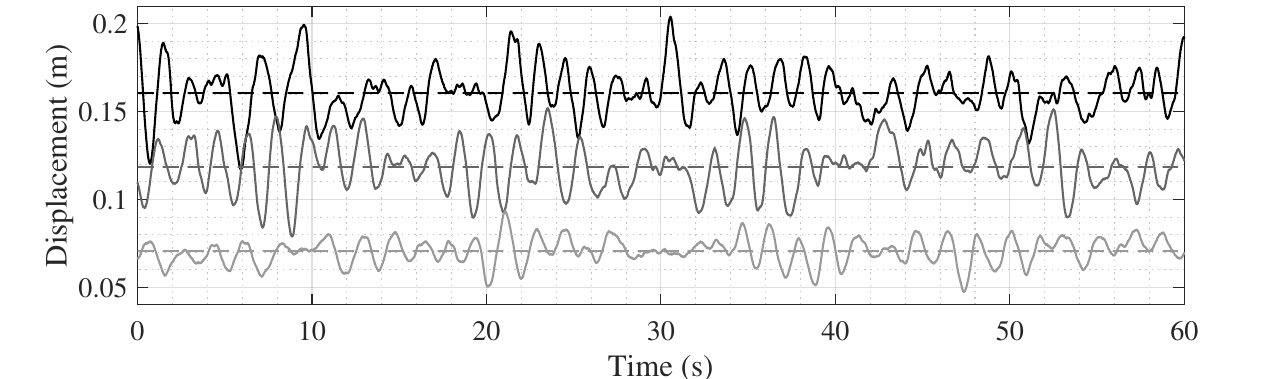}
    \caption{Examples of free end streamwise displacement vs.\ time over the 60 s steady-state periods for a PVC cylinder ($D = $ 3.8 $\pm$ 0.1 cm) for $U=$ [4.5, 5.6, 6.6] $\pm$ 0.5 ms$^{-1}$ (lines shaded from light to dark with increasing $U$). Mean displacements are indicated with dashed lines. The observed streamwise oscillations are consistent with previous studies in steady flow \citep{King1974VortexCurrents}}
    \label{fig:timetrace}
\end{figure}

The deflection for a given cylinder and wind speed was measured by tracking the free end of the cylinder in video frames collected at 240 frames per second (fps) with a resolution of 720 $\times$ 1280 pixels. The recording began with the cylinder at rest. The wind tunnel was then turned on and allowed to reach a steady-state speed. Measurements were collected for 60 seconds at each steady-state speed. The center of the cylinder surface was detected in each frame using a two-stage Hough Transform \citep{Yuen1990AFinding} in the MATLAB Image Processing Toolbox. The tip deflection, $\delta$, was determined by calculating the streamwise displacement of the center for each frame in the 60 s steady-state period with respect to its position under no wind load (figure \ref{fig:cylinder_displacement}). Examples of streamwise displacement versus time are shown in figure \ref{fig:timetrace}. Mean displacements were typically on the order of 250 pixels, corresponding to physical dimensions of approximately 10 cm. The normalized wind speed, $U/U_0$ was calculated by taking the mean of equation \ref{eq:ratio}. Note that while the cylinders were also free to move in the spanwise direction, spanwise displacements were an order of magnitude smaller than streamwise displacements and had a negligible mean value. Examples of cylinder free end trajectories showing both streamwise and spanwise displacements can be seen in the Supplementary Material (figure S1).

Six unique cylinders were tested. Cylinder aspect ratios ($L/D$) ranged from 30 to 48, and Reynolds numbers based on cylinder diameters ranged from 1.0 $\times$ $10^4$ to 2.2 $\times$ $10^4$. For rigid cylinders in this range of $Re_D$, the drag coefficient, $C_D$, should remain relatively constant, especially since the sharp drop in $C_D$ for cylinders is typically observed to occur at higher $Re_D$ ($\mathcal{O}(10^5)$) \citep{Roshko1961ExperimentsNumber}. Four of the cylinders were cut from lengths of soft, flexible PVC tubing (Masterkleer, Durometer 65A) with wall thickness $t=$ 0.6 $\pm$ 0.1 cm, length $L=$ 1.52 $\pm$ 0.01 m, and outer diameters $D \in$ [3.2, 5.1] $\pm$ 0.1 cm. The unprocessed PVC tubing tended to have some amount of curvature under no load. To achieve right cylindrical specimens, the PVC tubing was heated with boiling water and allowed to cool in a straightened position. To ensure that the effect of any remaining curvature leading to imperfect cylindrical shape did not systematically bias results, each PVC tube was tested three times, and rotated by $90^\circ$ about its axis in the mounting clamp in each subsequent test. 

\begin{table}[!ht]
  \begin{center}
\def~{\hphantom{0}}
\caption{Summary of test cylinder properties including the material, outer diameter, $D$, wall thickness of hollow tubes, $t$, length, $L$, aspect ratio, $L/D$, and Young's modulus, $E$. $^*$Note that for the PVC and polyurethane rubber cylinders, exact values of Young's modulus were not known, and  were therefore approximated based on the reported Shore hardness values using the relationship given in \cite{Gent1958OnModulus}.}
  \begin{tabular}{lccccc}
    \toprule
      Material & $D$ (cm) & $t$ (cm) & $L$ (m)  & $L/D$ & $E$ (MPa)\\
     \midrule
      PVC tube& 3.2 $\pm$ 0.1 & 0.6 $\pm$ 0.1 & 1.52 $\pm$ 0.01 & 48 & 4.4$^*$\\
      PVC tube& 3.8 $\pm$ 0.1 & 0.6 $\pm$ 0.1 & 1.52 $\pm$ 0.01 & 40 & 4.4$^*$\\
      PVC tube& 4.4 $\pm$ 0.1 & 0.6 $\pm$ 0.1 & 1.52 $\pm$ 0.01 & 35 & 4.4$^*$\\
      PVC tube& 5.1 $\pm$ 0.1 & 0.6 $\pm$ 0.1 & 1.52 $\pm$ 0.01 & 30 & 4.4$^*$\\
      Polyurethane Rubber (solid) & 3.2 $\pm$ 0.1 & N/A & 1.22 $\pm$ 0.01 & 38 & 1.7$^*$\\
      ABS tube with spring at base & 3.2 $\pm$ 0.1 & 0.3 $\pm$ 0.1 & 1.55 $\pm$ 0.01 & 48 & 2240 $\pm$ 380\\
      \botrule
  \end{tabular}
  \label{tab:cylinders}
  \end{center}
\end{table}

Two other cylinders of different types were tested in addition to the PVC tubes for the purpose of validating model applicability on cylinders with varying characteristics. One was a solid polyurethane rubber rod (medium soft, Durometer 40A) with dimensions $L = 1.22 \pm 0.01$ m and $D$ = 3.2 $\pm$ 0.1 cm. The final specimen was a hollow ABS tube ($L = 1.55$ $\pm$ 0.01 m, $D = 3.2$ $\pm$ 0.1 cm, $t =$ 0.3 $\pm$ 0.1 cm) with a steel compression spring at its base (spring rate of 2.8 N/mm) held by the mounting clamp at the tunnel roof. In this case, the ABS cylinder acted as a rigid body rather than a flexible cylinder like the other specimens, and the deflection was facilitated primarily by bending of the spring at its base. Table \ref{tab:cylinders} summarizes the properties of the six cylinders tested.

\subsection{Tree Deformation Measurements}
\label{sec:method_tree}

Measurements of two natural trees were conducted in a second, larger open-circuit wind tunnel of cross-section 2.88 m $\times$ 2.88 m. Deflections were measured for the two potted trees: a conifer commonly known as a juniper tree (\textit{Juniperus scopulorum}), and a broad-leaved bay laurel tree (\textit{Laurus nobilis}). Approximate tree dimensions are given in table \ref{tab:trees}. Trees were positioned using sand bags to weight down the pots and keep them in place. The tunnel was run in a uniform flow configuration at four distinct mean flow speeds spanning the tunnel range ($U =$ [3.3, 6.0, 8.8, 11.4] $\pm$ 0.5 ms$^{-1}$), which were measured using the factory-calibrated anemometer (Dwyer Series 641RM Air Velocity Transmitter). Reynolds numbers based on a length scale of $A^{1/2}$, where $A$ is the approximate frontal area of the tree under no load, ranged from $1.7 \times 10^5$ to $6.3 \times 10^5$. The maximum blockage ratio based on projected frontal area was 8.4$\%$.

\begin{table}[!hbt]
  \begin{center}
\def~{\hphantom{0}}
  \caption{Tree properties including tree height, $L$, measured from base to tip, frontal area, $A$, and trunk diameter at breast height, $DBH$. Approximate heights and areas were measured using photos taken from downstream of the trees under no wind load. Note values of $A$ were estimated using the outer envelope of the trees, and do not incorporate leaf density.}
  \begin{tabular}{lccc}
    \toprule
  & \textit{Juniperus scopulorum}& \textit{Laurus nobilis}\\
      & (Juniper) & (Bay Laurel)\\
         \midrule
      $L$ (m) & 1.6 $\pm$ 0.1 & 1.7 $\pm$ 0.1 \\
      $A$ (m$^2$)& 0.6 $\pm$ 0.1 & 0.7 $\pm$ 0.1\\
      $DBH$ (cm) & 0.7 $\pm$ 0.1 & 3.0 $\pm$ 0.1\\
    \botrule
  \end{tabular}
  \label{tab:trees}
  \end{center}
\end{table}

\begin{figure}[!hbt]
    \centering
    \includegraphics[width = \linewidth, trim=110 260 110 60, clip]{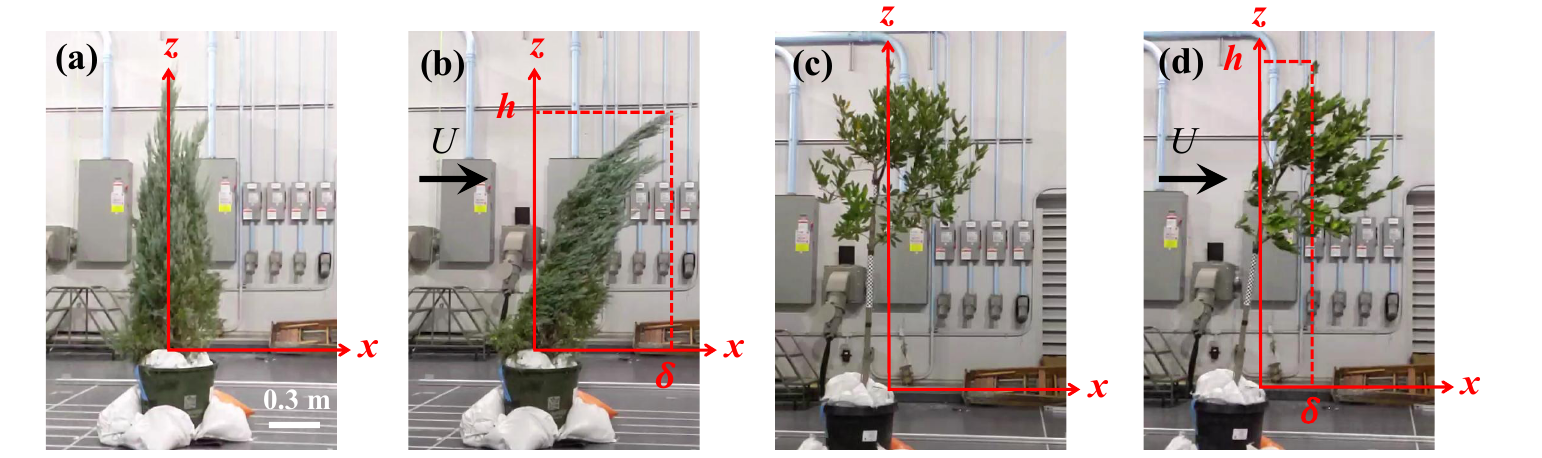}
    \caption{Example measurements of streamwise deflection, $\delta$, and projected height, $h$, for the juniper tree (a, b) and the laurel tree (c, d). Measurements were made in reference to treetop position under no wind load (a, c). Resulting measurements of tree deformation for $U=$ 11.4 ms$^{-1}$ are shown in (b, d)}
    \label{fig:tree_measurements}
\end{figure}

The camera was angled perpendicular to the flow direction to capture video of the streamwise tree deflections for 110 s at each steady-state flow speed. Videos were recorded at 240 fps with a resolution of 720 $\times$ 1280 pixels. A random sample of frames was chosen to measure the tip deflections. For each 10-second clip of video data, 10 frames were randomly selected. This yielded 110 sample frames for each flow speed. Values for $\delta$ and $h$ were measured over the 110 samples. The pixel position of highest point on the tree was manually identified in each sample frame using MATLAB. This position was used to calculate $h$ and $\delta$ for each frame. The streamwise position of the highest point on the tree under no wind load marked $\delta = 0$, and $h$ was measured in reference to the base of the tree. Representative measurements of $h$ and $\delta$ are shown in figure \ref{fig:tree_measurements}. The values of $h$ and $\delta$ found for each frame were used to calculate $U/U_0$ as shown in equation \ref{eq:ratio_trees}. The mean value of was taken over the 110 sample frames for each flow speed.

\section{Results}

\subsection{Normalized Wind Speeds from Cylinder Deflections}
\label{sec:results_cylinders}

\begin{figure}[!hbt]
    \centering
    \includegraphics[width = 0.85\linewidth, trim=0 200 0 200, clip]{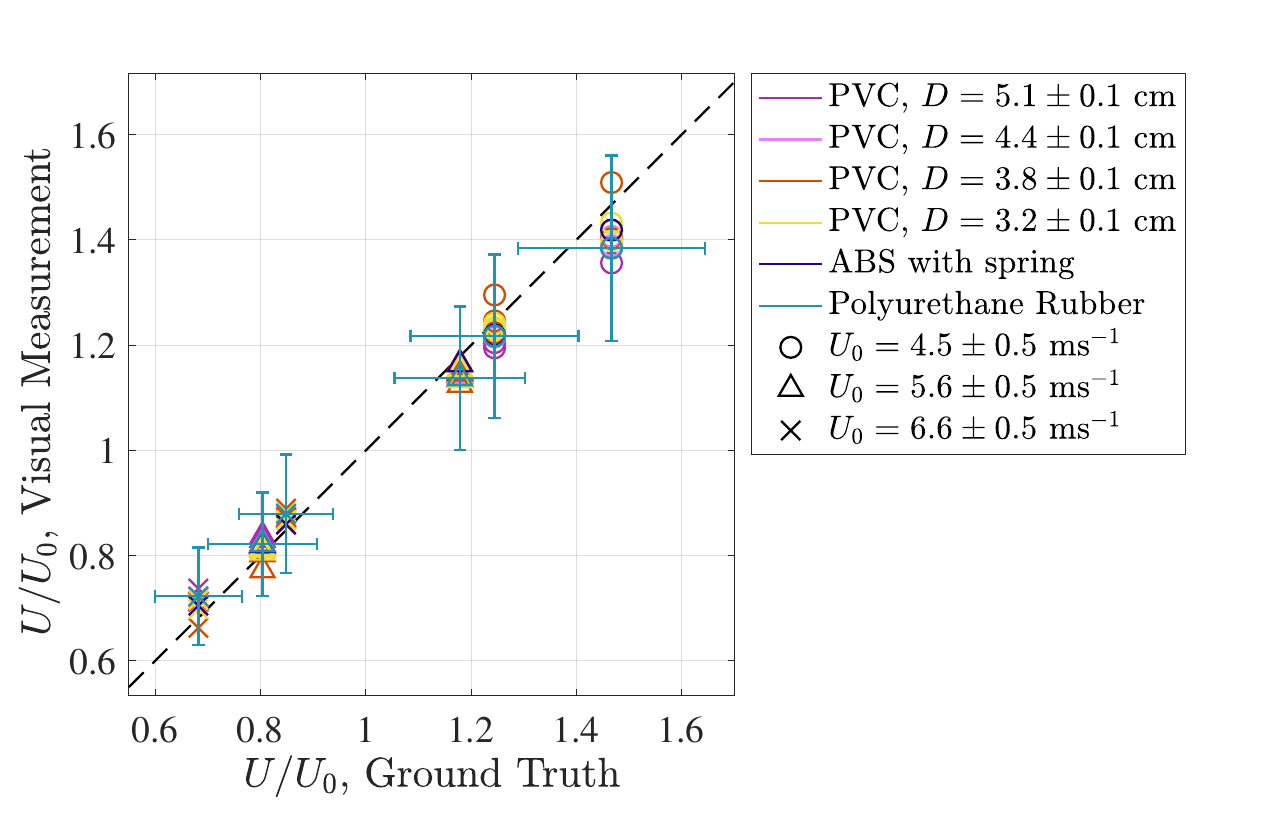}
    \caption{Visually measured normalized wind speed vs.\ ground truth for all test cylinders. Marker colors indicate unique cylinders, and marker types indicate the reference speed, $U_0$. The dashed black line represents unity}
    \label{fig:cylinder_results}
\end{figure}

Normalized wind speeds inferred from visual measurements and equation \ref{eq:ratio} are compared to ground truth anemometer-measured values in figure \ref{fig:cylinder_results}. Results are shown for all six test cylinders. Each datapoint represents $U/U_0$ calculated using equation \ref{eq:ratio} for a particular test cylinder based on deflections $\delta$ and $\delta_0$ at wind speeds $U$ and $U_0$ respectively. As discussed in section \ref{sec:methods_cylinder}, one reference condition ($U_0$, $\delta_0$) must be known to calculate $U/U_0$. The results shown in figure \ref{fig:cylinder_results} include datapoints calculated using each of the three distinct mean flow speeds ($U=$ [4.5, 5.6, 6.6] $\pm$ 0.5 ms$^{-1}$) as the reference, $U_0$, to determine $U/U_0$ for the other two tunnel speeds. Thus, using each combination of two wind speeds to obtain $U/U_0$ yields six ground truth values for which visually measured results can be compared (two datapoints for each distinct reference wind speed). Distinct markers are used to denote datapoints calculated with each particular value of $U_0$. Representative error bars are shown for the cylinder with the largest overall error values (the polyurethane rubber cylinder). The error bars were of the same order of magnitude for all cylinders. Horizontal error bars were calculated by propagating the uncertainty from anemometer accuracy through the ratio $U/U_0$. Vertical error bars were calculated based on the standard deviation of deflections over the tracking period, which were propagated through equation \ref{eq:ratio}. The dashed black line shows the exact one-to-one relationship that would indicate perfect agreement between visually measured speeds and ground truth values. There is strong agreement between the visual and ground truth measurements. 

\subsection{Normalized Wind Speeds from Tree Deflections}
\label{sec:results_trees}

\begin{figure}[!hbt]
    \centering
    \includegraphics[width = \linewidth, trim=0 210 0 180, clip]{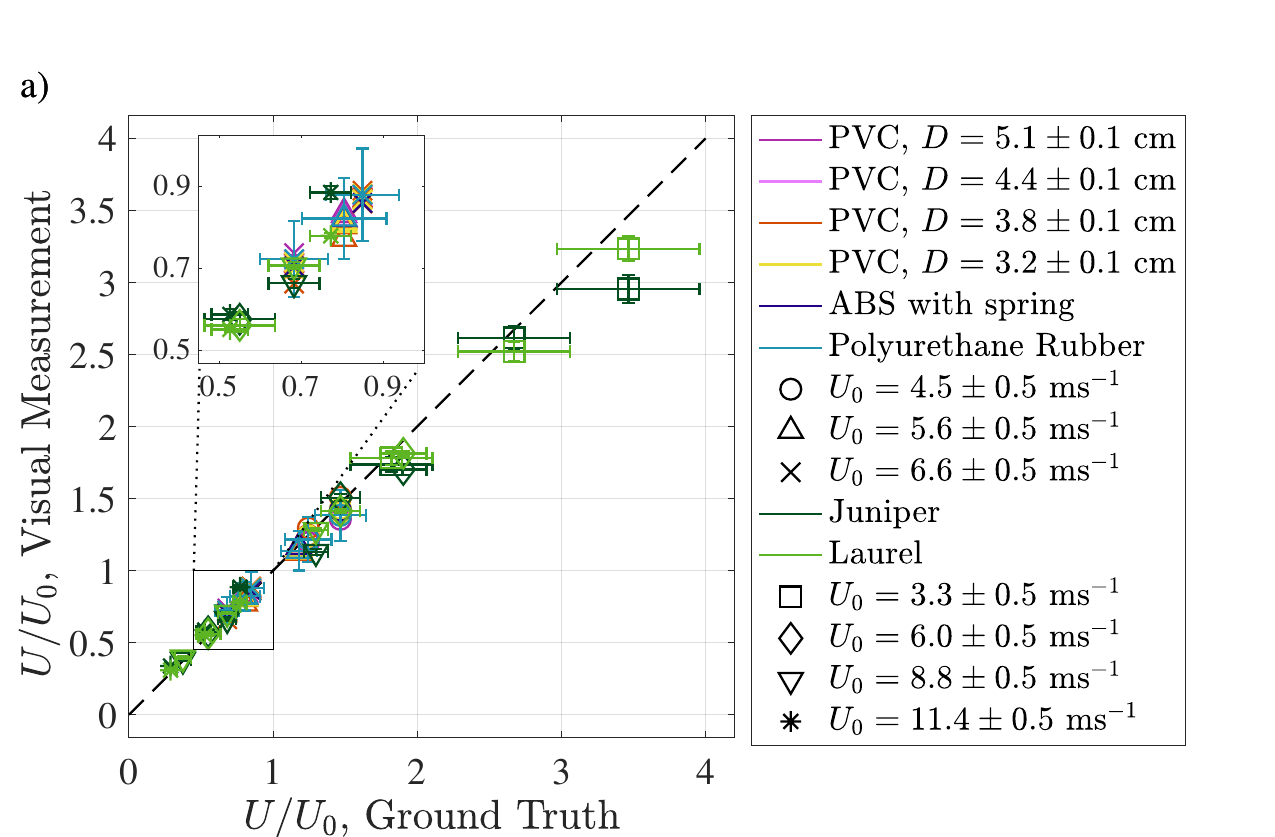}
    \includegraphics[width = \linewidth, trim=0 210 0 180, clip]{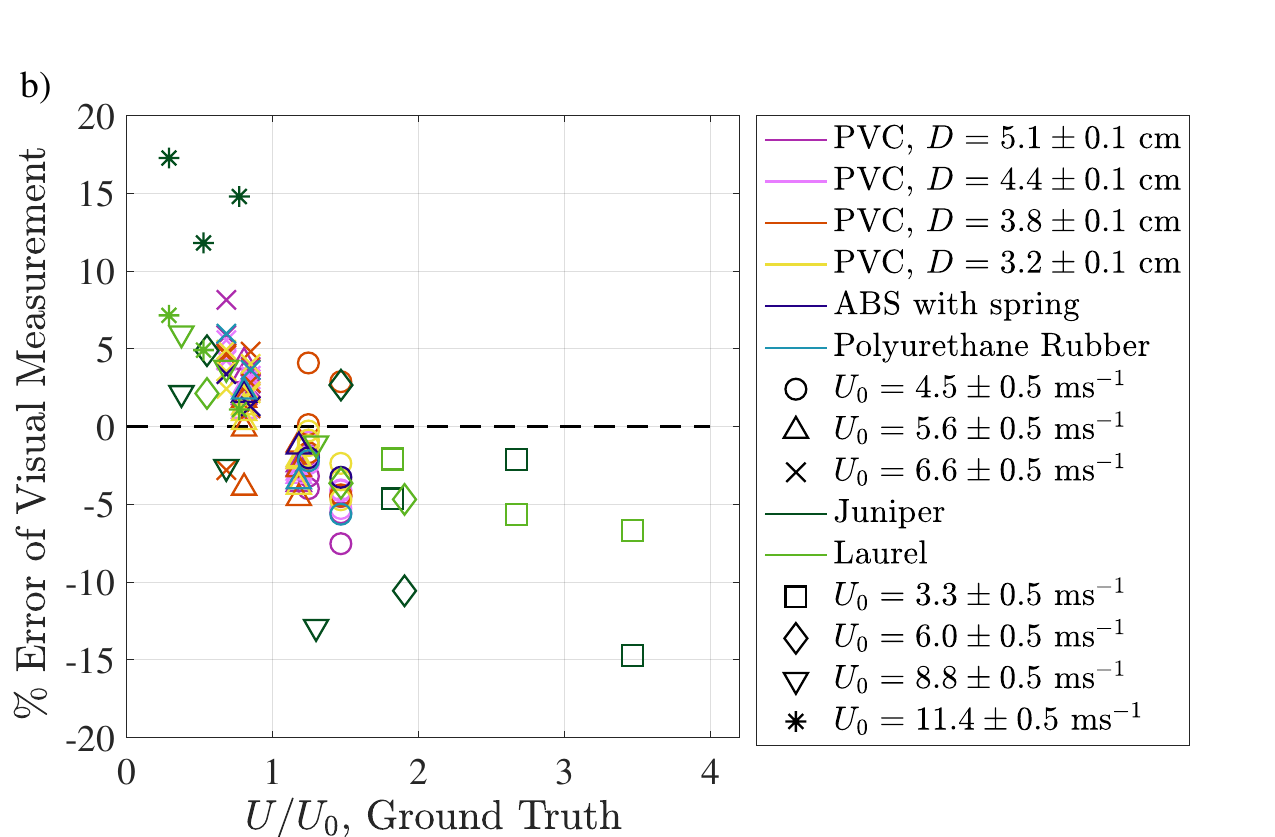}
\caption{(a): Visually measured normalized wind speed vs.\ ground truth for trees and cylinders. Marker colors indicate the sample object, and marker types indicate the reference wind speed used in the visual measurements. The dashed black line represents unity indicating perfect model agreement. (b): Percent error of visual measurement compared to ground truth}
    \label{fig:tree_results}
\end{figure}

\begin{figure}[!hbt]
    \centering
    \includegraphics[width = \linewidth, trim=0 210 0 180, clip]{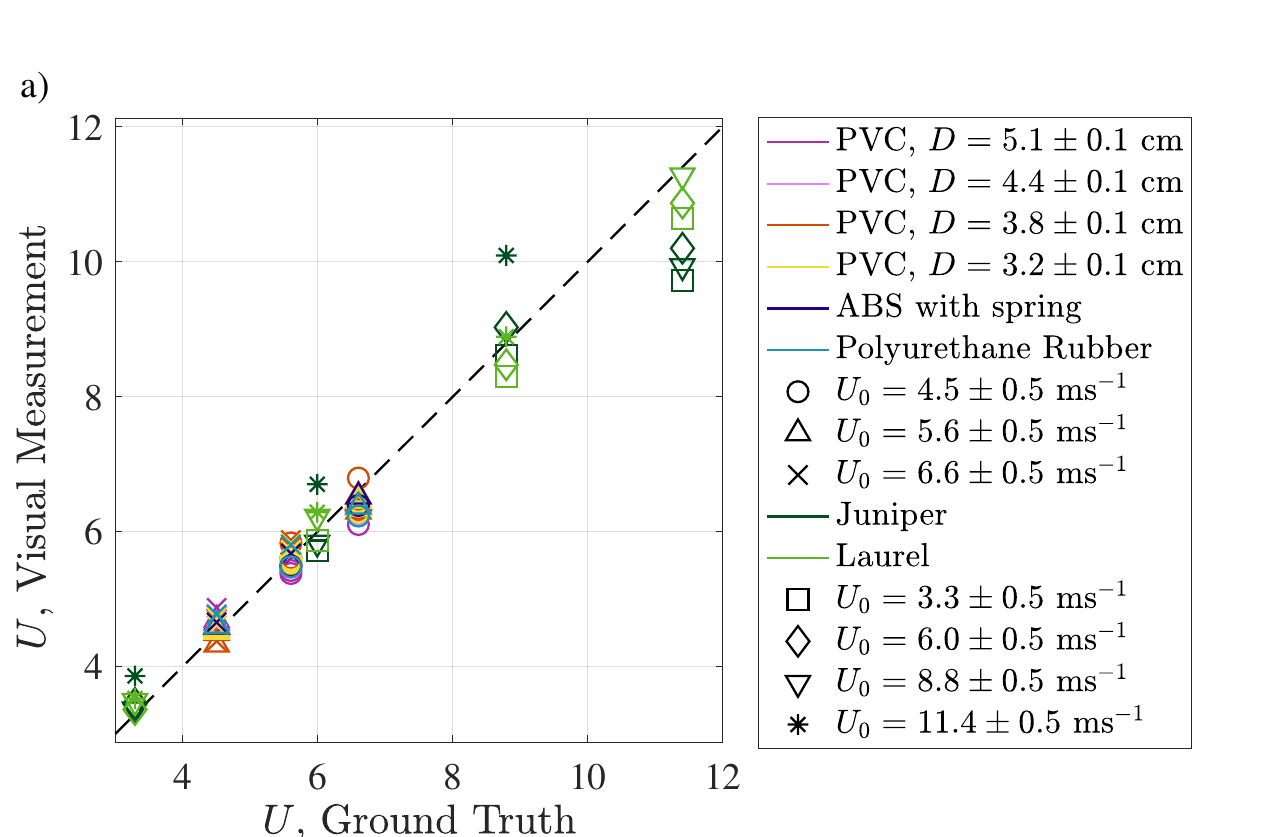}
    \includegraphics[width = \linewidth, trim=0 210 0 180, clip]{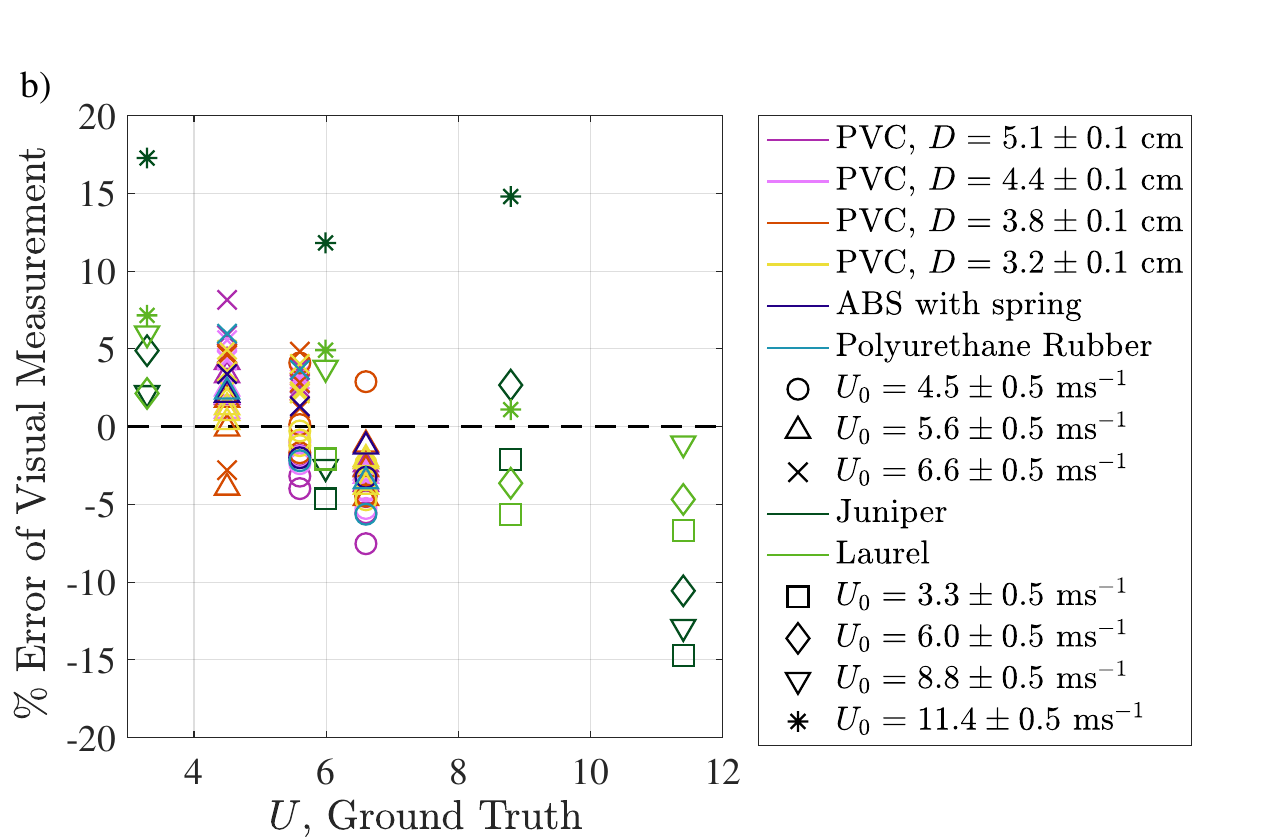}
\caption{(a): Visually measured dimensional wind speed vs.\ ground truth for trees and cylinders. Marker colors indicate the sample object, and marker types indicate the reference wind speed used in the visual measurements. The dashed black line represents unity indicating perfect model agreement. (b): Percent error of the dimensional visual measurement compared to ground truth}
    \label{fig:tree_results_dimensional}
\end{figure}

Visually measured normalized wind speeds versus ground truth normalized wind speeds are plotted in figure \ref{fig:tree_results}a for all test objects including the two trees and six cylinders. Note that cylinder datapoints are the same as they appear in figure \ref{fig:cylinder_results}, but are also included in figure \ref{fig:tree_results}a for ease of comparison between cylinder and tree results. The model accounting for changing frontal area (equation \ref{eq:ratio_trees}) was used to calculate the visually measured normalized wind speeds for the trees. Each of the four tunnel speeds used in the tree experiments ($U =$ [3.3, 6.0, 8.8, 11.4] $\pm$ 0.5 ms$^{-1}$), was taken as a reference speed $U_0$ in combination with the other three distinct speeds for a total of 12 ground truth values. Once again, distinct markers have been used to show points calculated with each particular reference speed. The trees were subjected to a wider range of wind speeds than the cylinders based on tunnel capabilities, which led to results across a wider range of ground truth values ($U/U_0 \in [0.3, 3.5]$ for the trees compared to $U/U_0 \in [0.7, 1.5]$ for the cylinders), as visible in figure \ref{fig:tree_results}a. The percent error of visual measurements compared to ground truth is shown in figure \ref{fig:tree_results}b. Dimensional wind speeds recovered by multiplying the visually measured normalized speed by the reference speed, $U_0$, are shown in figure \ref{fig:tree_results_dimensional}. The accuracy of the visual measurement approach is validated in the comparison with ground truth. The visual measurements appear to agree well with ground truth measurements, but do modestly underestimate the ground truth at high wind speeds. This could be associated with the limitation in our estimate of area change, as we assumed $A \propto h^2$ to incorporate the tree crown deformation given only a side view of the tree. In addition, the breakdown of the model at large values of $U/U_0$ suggests a potential departure from the assumptions of the linear model utilized in computing the wind speed. We examine this latter possibility in more detail in the following section.

\section{Discussion and Conclusions}
\label{sec:discussion}

Wind speed measurements inferred from structure kinematics showed strong agreement with the ground truth measurements for both the cylinders and the trees, as seen in figures \ref{fig:cylinder_results} and \ref{fig:tree_results}. This suggests that flow-structure interactions can be leveraged to infer local wind conditions without \textit{a priori} knowledge of the material properties of the structure, and without visualizing or instrumenting the flow itself.

\begin{figure}[!h]
    \centering
    \includegraphics[width = 0.9\linewidth, , trim=0 140 0 170, clip]{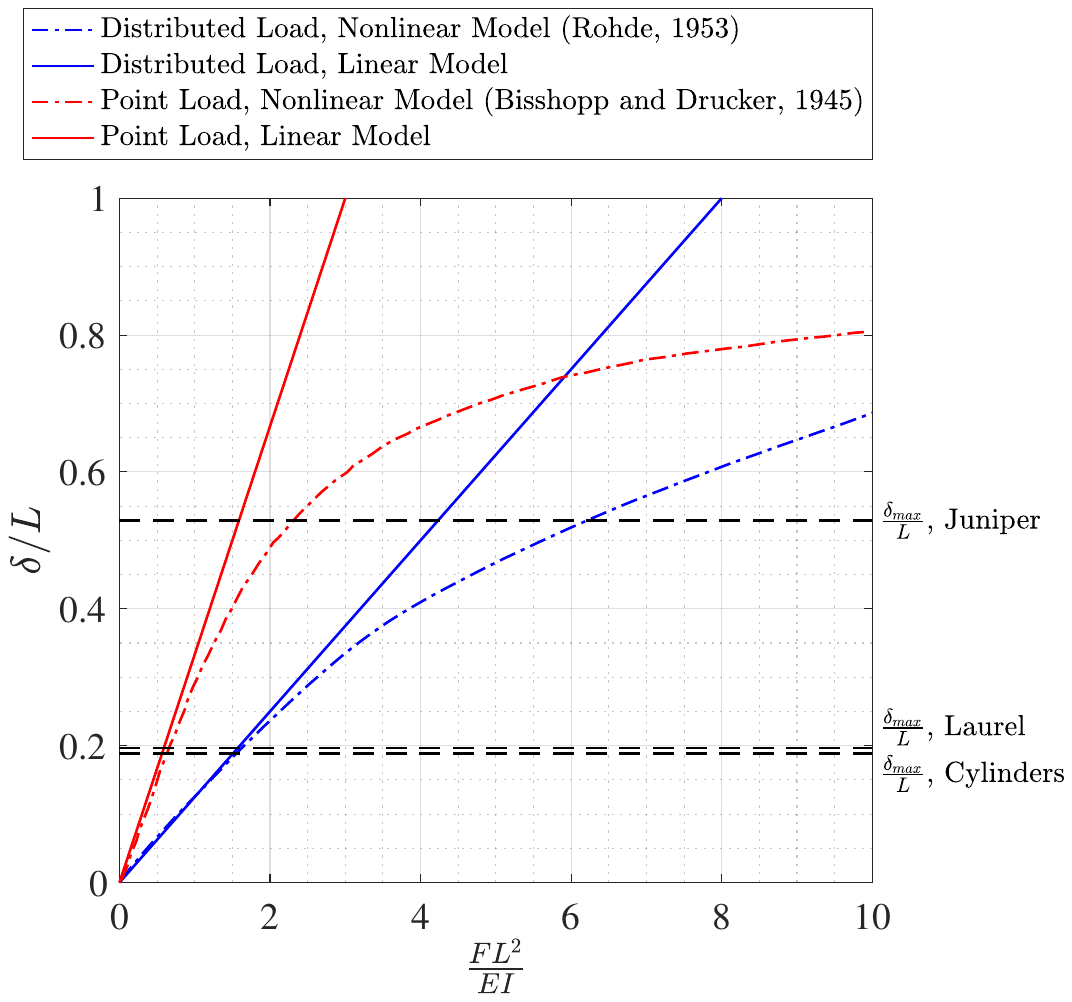}
    \caption{Normalized deflection ($\delta / L$) of a cantilever beam vs. normalized force ($\frac{FL^2}{EI}$). Curves are given for nonlinear models for a uniformly distributed load and for a point load reproduced from \citet{Rohde1953LargeLoad} and \citet{Bisshopp1945LargeBeams} respectively, shown along with the linear relationships given by elementary theory. The maximum deflection observed in the present work for the juniper tree, laurel tree, and cylinders are shown with dashed black lines}
    \label{fig:large_bending}
\end{figure}

The agreement in normalized wind speed is particularly good at lower values of $U/U_0$, but there is more discrepancy at higher values. As seen in figure \ref{fig:tree_results}b, the percent error in the visually measured normalized speed increases in magnitude as the ground truth normalized speed departs from a value of 1. One factor that may contribute the discrepancy is that the model assumptions (e.g.\ the linear assumption of Hooke's law) become less valid at higher wind speeds where there are larger forces acting on the structures causing larger deflections. For instance, assuming a linear relationship for a cantilever beam subject to a uniformly distributed load or a point load overestimates the tip deflection when deformations are large \citep{Rohde1953LargeLoad, Bisshopp1945LargeBeams}. This linear model would then underestimate the wind speed corresponding to a given deflection. This is reflected in the overestimates for ground truth values of $U/U_0<1$ and the underestimates for ground truth values of $U/U_0>1$ in figure \ref{fig:tree_results}b. Figure \ref{fig:large_bending} shows the deflections expected using nonlinear beam bending models for both a uniformly distributed load \citep{Rohde1953LargeLoad} and a point load \citep{Bisshopp1945LargeBeams}. 

\noindent For both loading conditions, the nonlinear bending models start to noticeably depart from the linear models at deflections greater than $\delta / L \approx 0.2$. Although the deflections observed for the cylinders and the laurel tree typically fell within the linear regime, the maximum deflections observed for the juniper tree were large enough to expect a significant departure. Notably, the juniper tree also yielded the highest error in visual wind speed measurements of the objects tested, particularly for datapoints that relied on deflections at the highest wind speed of 11.4 ms$^{-1}$ (figure \ref{fig:tree_results}b).

\begin{figure}[!h]
    \centering
    \includegraphics[width = 0.9\linewidth, trim=0 210 0 210, clip]{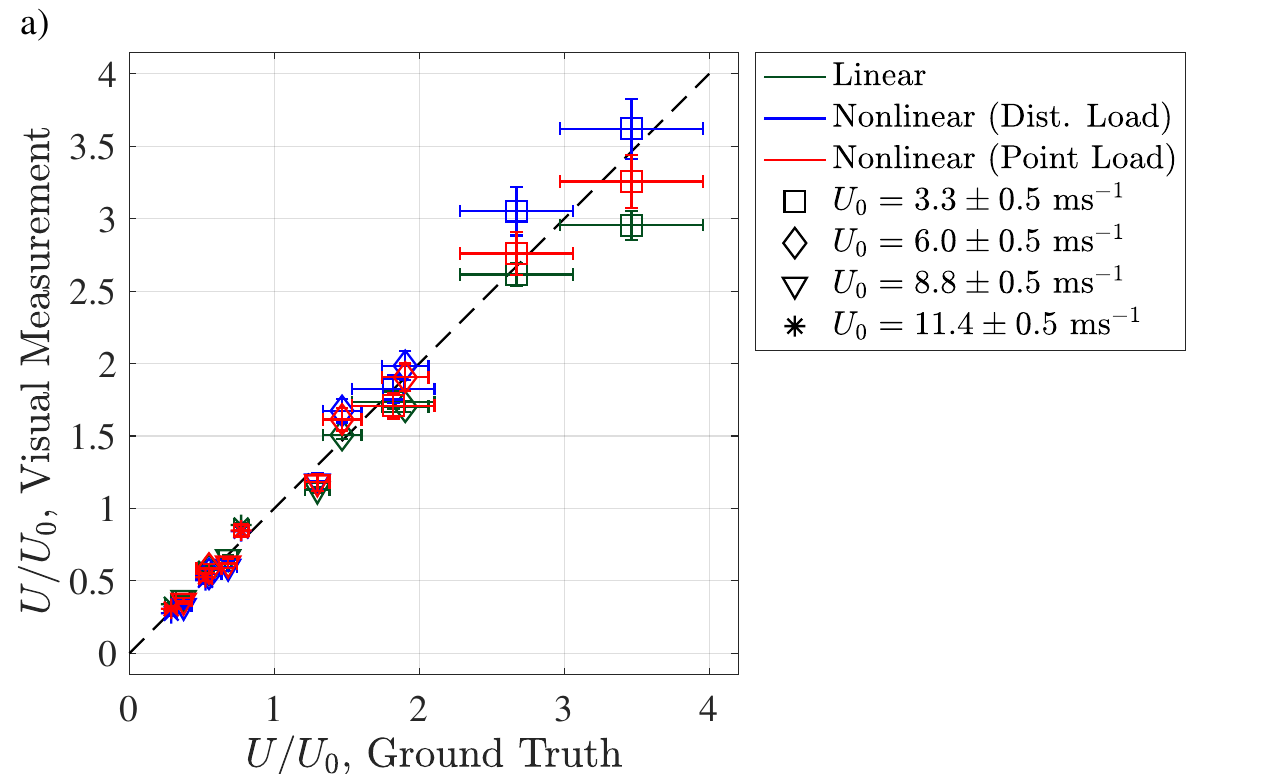}
    \includegraphics[width = 0.9\linewidth, trim=0 210 0 210, clip]{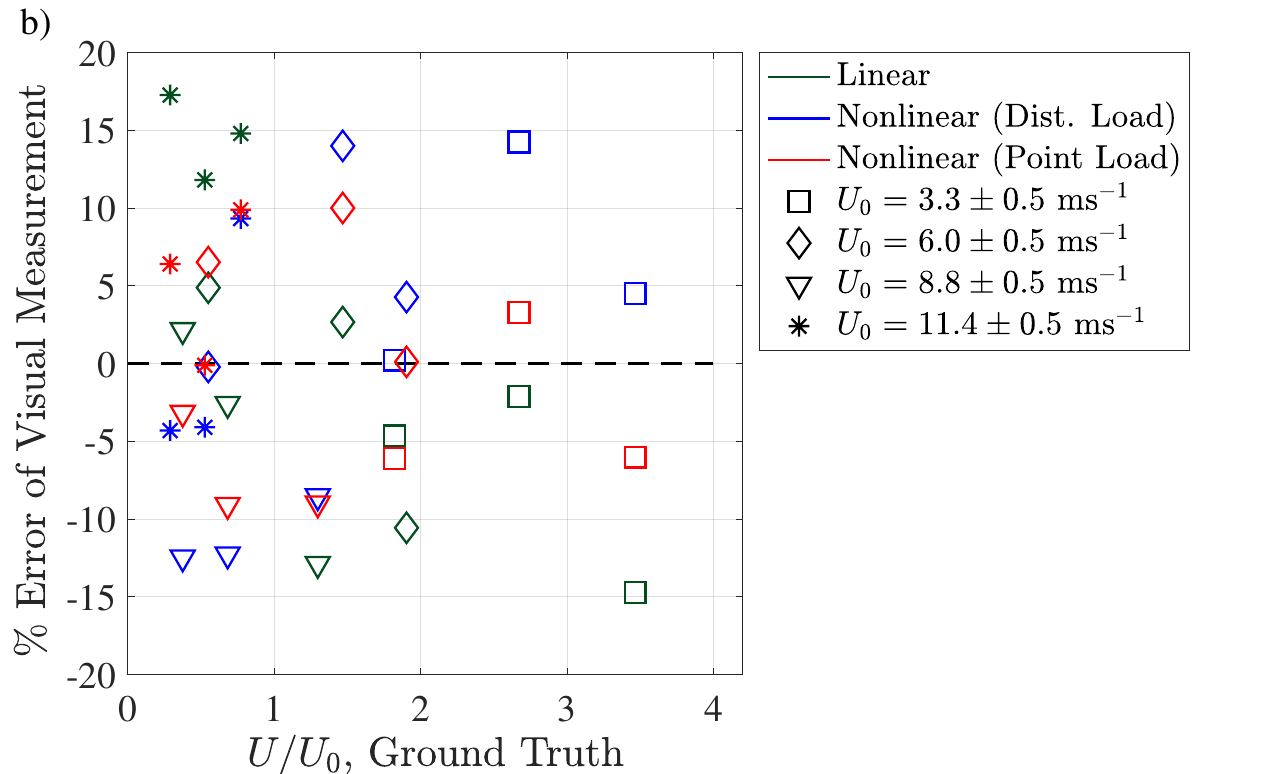}
\caption{(a): Visually measured normalized wind speed vs.\ ground truth for the juniper tree using original linear model as well as models corrected according to the nonlinear models for beam bending under a point load \citep{Bisshopp1945LargeBeams} and distributed load \citep{Rohde1953LargeLoad}. (b): Percent error of visual measurement compared to ground truth}
    \label{fig:large_bending_correction}
\end{figure}

To further consider the effects of large deflections of the juniper tree, the nonlinear models discussed in \citet{Rohde1953LargeLoad} and \citet{Bisshopp1945LargeBeams}  were used to correct the visual wind speed measurements. Note that the nonlinear models incorporate additional assumptions regarding the loading configuration, i.e. a point load at the tip \citep{Bisshopp1945LargeBeams} or a uniformly distributed load \citep{Rohde1953LargeLoad}. While neither of these models strictly captures the real tree loading, they provide a means to conceptually assess the effect of nonlinearity. Cubic spline interpolation was used to obtain numerical functions for the nonlinear bending models from the curves shown in figure \ref{fig:large_bending}. Observed values of $\delta / L$ for the juniper tree were used to determine the expected normalized force according to the nonlinear models. This was used to find the effective deflection, $\hat{\delta}$, that would have been expected using the linear models with the same given force. Finally, the values of $\hat{\delta}$ were used to obtain the normalized wind speed measurements from equation \ref{eq:ratio_trees}. The results of these corrections for large bending are shown in figure \ref{fig:large_bending_correction}. While the error in visual measurements is not eliminated, the maximum error is reduced. For the linear model, the magnitude of error generally increased as larger wind speeds were involved in the calculation and $U/U_0$ departed from a value of 1. This trend is no longer apparent after the corrections for large bending have been applied \ref{fig:large_bending_correction}b. This suggests that the nonlinearity not captured by the original model likely plays a role in the increased error at higher wind speeds. It is possible that the application of other beam bending models meant for large deformations may improve agreement. Prior works have successfully described the bending behavior of sapling trunks using tapered cantilevered beams models for analyzing large deflections \citep{Kemper1968LargeBeams, Morgan1987StructuralLoading, Gardiner1992MathematicalTrees}. However, these more complex models required knowledge the flexural rigidity and the taper of the trunk, both of which may be unknown.

Future studies may consider the use of a second camera aligned with the direction of flow to capture the instantaneous frontal area of the trees. This would enable validation of the model assumption $A \propto h^2$ applied in equation \ref{eq:ratio_trees}. However, in practice, for many potential field applications it is beneficial to use only a single camera, as this reduces the complexity of the setup and does not require access to multiple spatial locations surrounding the object of interest.

In the present work, the flow incident on the structures is designed to be spatially uniform. However, wind speeds in the near-surface region of the atmospheric boundary layer can vary with height above the ground, $z$. A common model is the log wind profile:

\begin{equation}
    U(z) = \frac{u^*}{K}\ln \left( \frac{z + d}{z_0} \right)
    \label{eq:abl}
\end{equation}

\noindent where $u^*$ is the friction velocity, $K$ is the von K\'{a}rm\'{a}n constant ($K \approx 0.4$), $d$ is the displacement height, and $z_0$ is the surface roughness \citep{Stull1988AnMeteorology}. The model presented here is still applicable for the loading on a cantilever beam resulting from the wind profile in equation \ref{eq:abl}. Under linear beam theory, the equation for tip displacement, $\delta$, is a linear differential equation. Hence, the principle of superposition applies, and the deflection from an arbitrary distributed load can be determined by superposing the solutions for the differential load at each point along the length of the beam \citep{Bauchau2009StructuralAnalysis}. The tip deflection due to a point load, $P$, acting at a height of $z$ along the beam is given by:

\begin{equation}
\delta = \frac{P z^2 (3L - z)}{6EI}
\end{equation}

\noindent Thus, for a distributed force per unit length, $f(z)$, the resulting tip deflection can be calculated as:

\begin{equation}
    \delta =  \frac{1}{6EI} \int_{0}^{L} f z^2 (3L - z)dz
\end{equation}

\noindent Since the force from the wind is proportional to the square of the wind speed (equation \ref{eq:F_D}), the distributed load due to the log wind profile is given by $f(z) = c U^2(z) $, where $c$ is a constant, resulting in the tip deflection:

\begin{equation}
    \delta = \frac{c \left(\frac{u^*}{K}\right)^2}{6EI} \int_{d}^{L}  \ln^2 \left( \frac{z + d}{z_0} \right) z^2 (3L - z)dz
    \label{eq:delta_abl}
\end{equation}

\noindent If $z_0$ and $d$ remain constant for the conditions of interest (i.e. the shape of the $U(z)$ remains the same), then it follows from equation \ref{eq:abl} that $u^* \propto U(L)$. In this case, since $u^*$ is the only component of \ref{eq:delta_abl} that changes with the mean wind speed, the relationship $U(L) \propto \sqrt{\delta}$ applies, and thus, the formulation of the model given in equation \ref{eq:ratio} can still be used to calculate the normalized wind speeds. In general, this approach of calculating $\delta$ from linear beam theory with a distributed load resulting from a particular wind profile will show that equation \ref{eq:ratio} is applicable if the shape of $U(z)$ remains the same between the reference and desired measurement conditions, with its magnitude scaling with the mean wind speed at the height of interest. This does not take into account the tree crown deformation, which may also vary with $z$. Future studies may consider this effect, as well as the influence of diurnal and seasonal variation of the velocity profile in the atmospheric boundary layer.

A further limitation of the present work is the need for a calibration reference ($U_0$, $\delta_0$) to convert the normalized wind speeds to dimensional quantities. While the dimensional wind speeds are needed to determine other dimensional quantities such as kinetic energy flux, the normalized measurements may be sufficient for determining other useful flow characteristics even without calibration. For example, the shape factor, $k$, of the probability density function of a Weibull distribution of wind speeds can be determined using the normalized wind speed measurements. Weibull distributions are often used to characterize wind speed data \citep{Takle1977NoteData} and are useful for applications such as assessing sites for wind energy generation \citep{Justus1976NationwideGenerators}. The maximum likelihood method is commonly used to determine the Weibull distribution parameters \citep{Seguro2000ModernAnalysis}, with $k$ estimated by iteratively solving the implicit equation:

\begin{equation}
    k = \left( \frac{\sum_{i=1}^n{ \left(\frac{U_i}{U_0}\right) ^k\ln \left(\frac{U_i}{U_0}\right)}} {\sum_{i=1}^n{\left(\frac{U_i}{U_0}\right)^k}} - \frac{\sum_{i=1}^n{\ln \left(\frac{U_i}{U_0}\right)}} {n} \right)^{-1}
    \label{eq:shape}
\end{equation}

\noindent where $\frac{U_i}{U_0}$ are the normalized wind speed measurements. 

Future work will seek to combine the present flow physics-based approach with data-driven approaches to approximate structural parameters necessary to provide the calibration reference solely from visual measurements of the structures in the flow. Moreover, while the present work focused on trees as natural, visual anemometers, the concept can also be extended to other objects that are prevalent in important environmental flows, such as flags in the built environment \citep{Cardona2019SeeingNetwork} and seagrass in ocean currents \citep{Zeller2014ImprovedExperiments}.

\begin{Backmatter}

\paragraph{Acknowledgments}
The authors would like to thank Peter Gunnarson, Berthy Feng, and Emily de Jong for their assistance in running wind tunnel experiments, and Matthew Fu for his thoughtful comments and discussion.

\paragraph{Funding Statement}
This work was supported by the National Science Foundation (grant CBET-2019712), and by the Center for Autonomous Systems and Technologies at Caltech.

\paragraph{Competing Interests}
The authors report no conflict of interest.

\paragraph{Data Availability Statement}
The data discussed in this work will be made available at the Stanford Digital Repository at \href{https://purl.stanford.edu/tp480sx4819}{https://purl.stanford.edu/tp480sx4819}.  

\paragraph{Author Contributions}
Conceptualization: JLC; KLB; JOD. Methodology: JLC; JOD. Investigation: JLC. Software: JLC. Data analysis: JLC; JOD. Funding acquisition: KLB; JOD.

\paragraph{Supplementary Material}
Additional information can be found in the supplementary material.

\bibliographystyle{apalike}
\bibliography{references.bib}

\end{Backmatter}

\end{document}


\begin{Frontmatter}

\title[]
{Supplementary Material}

\author[1]{Jennifer L. Cardona}\orcid{0000-0003-3491-0638}
\author[2]{Katherine L. Bouman}\orcid{0000-0003-0077-4367}
\author*[3]{John O. Dabiri}\email{jodabiri@caltech.edu}


\address[1]{\orgdiv{Department of Mechanical Engineering}, \orgname{Stanford University}, \orgaddress{\street{Stanford}, \state{California}, \postcode{94305}, \country{USA}}}

\address*[2]{\orgdiv{Computing and Mathematical Sciences \& Electrical Engineering \& Astronomy}, \orgname{California Institute of Technology}, \orgaddress{\street{Pasadena}, \state{California}, \postcode{91125}, \country{USA}}}

\address*[3]{\orgdiv{Graduate Aerospace Laboratories \& Mechanical Engineering}, \orgname{California Institute of Technology}, \orgaddress{\street{Pasadena}, \state{California}, \postcode{91125}, \country{USA}}}

\received{}
\revised{}
\accepted{}

\end{Frontmatter}
\newcommand{\beginsupplement}{%
        \setcounter{table}{0}
        \renewcommand{\thetable}{S\arabic{table}}%
        \setcounter{figure}{0}
        \renewcommand{\thefigure}{S\arabic{figure}}%
     }
\beginsupplement

\section{Cylinder Trajectories}
\label{sec:trajectories}
\begin{figure}[!ht]
    \centering
    \includegraphics[width = 0.32\linewidth, trim=135 200 150 200, clip]{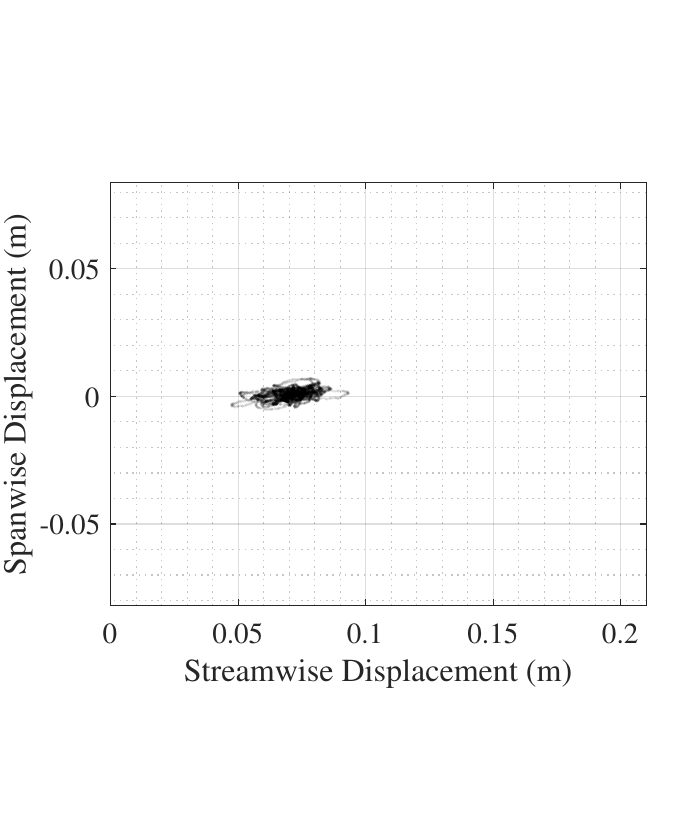}
    \includegraphics[width = 0.32\linewidth, trim=135 200 150 200, clip]{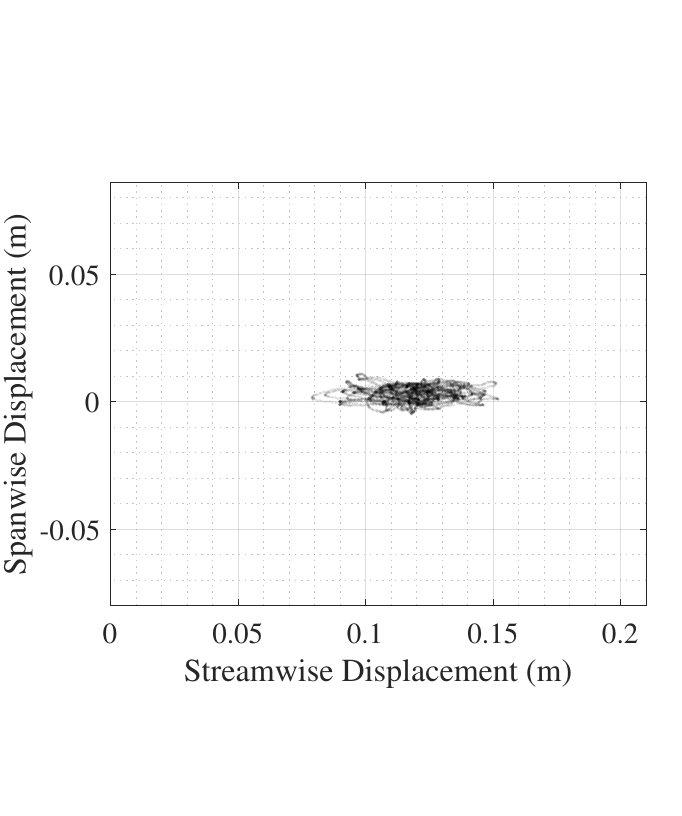}
    \includegraphics[width = 0.32\linewidth, trim=135 200 150 200, clip]{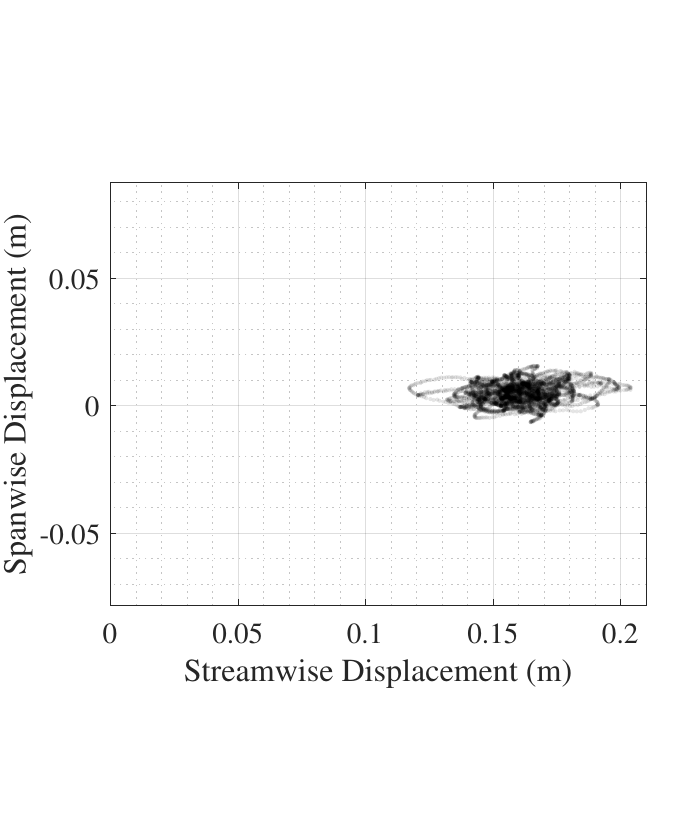}
    \caption{Example trajectories for the free end of a cylinder (PVC tube, $D =$ 3.8 $\pm$ 0.1 cm). Each point corresponds the position of the cylinder free end at one instant in time subject to flow speeds $U =$ 4.5 $\pm$ 0.5 ms$^{-1}$ (left), $U =$ 5.6 $\pm$ 0.5 ms$^{-1}$ (center), and $U =$ 6.6 $\pm$ 0.5 ms$^{-1}$ (right) over the 60 s steady state period for each speed. Note that while the data was collected at 240 Hz, the data shown here have been down-sampled to 80 Hz to reduce overlapping points. All axes are identically scaled}
    \label{fig:trajectories}
\end{figure}

Figure \ref{fig:trajectories} shows examples of cylinder free end trajectories. Notably, streamwise displacement values are an order of magnitude larger than spanwise displacements at all points in time, and streamwise displacements have a non-zero mean.

